\documentclass[pra,reprint,aps,twocolumn,showpacs,superscriptaddress,floatfix,amsmath,amssymb,nofootinbib]{revtex4-2}

\usepackage{graphicx}

\usepackage{float}
\usepackage{titlesec}
\titlespacing*{\section}{0pt}{10pt}{10pt} 
\usepackage{dcolumn}
\usepackage{bm}
\usepackage{hyperref}
\usepackage{xcolor}
\newcommand{\ketbra}[2]{\ensuremath{\ket{#1}\!\bra{#2}}}
\newcommand{\beq}{\begin{equation}}
\newcommand{\eeq}{\end{equation}}
\newcommand{\beqa}{\begin{eqnarray}}
\newcommand{\eeqa}{\end{eqnarray}}

\usepackage{physics}
\usepackage{comment}
\usepackage{ulem}

\newcommand{\appropto}{\mathrel{\vcenter{
 \offinterlineskip\halign{\hfil$##$\cr
  \propto\cr\noalign{\kern2pt}\sim\cr\noalign{\kern-2pt}}}}}

\hypersetup{colorlinks,citecolor=blue,filecolor=blue,linkcolor=cyan,urlcolor=blue}

\begin{document}

\title[Characterization of the PES in NV centers through quantum jumps]{Characterization of the photon emission statistics in nitrogen-vacancy centers} 

\author{I. Panadero}
\affiliation{Arquimea Research Center, Camino las Mantecas s/n, 38320 San Cristobal de La Laguna, Spain}
\affiliation{Department of Physics, Universidad Carlos III de Madrid, Avda. de la Universidad 30, 28911 Legan\'es, Spain}
\affiliation{Department of Physical Chemistry, University of the Basque Country UPV/EHU, 48080 Bilbao, Spain}
\author{H. Espin\'os}
\affiliation{Department of Physics, Universidad Carlos III de Madrid, Avda. de la Universidad 30, 28911 Legan\'es, Spain}
\author{L. Tsunaki}
\affiliation{ Helmholtz-Zentrum Berlin f\"ur Materialien und Energie GmbH, Hahn-Meitner-Platz 1, 14109 Berlin, Germany}
\author{K. Volkova}
\affiliation{ Helmholtz-Zentrum Berlin f\"ur Materialien und Energie GmbH, Hahn-Meitner-Platz 1, 14109 Berlin, Germany}
\author{A. Tobalina}
\affiliation{Arquimea Research Center, Camino las Mantecas s/n, 38320 San Cristobal de La Laguna, Spain}
\author{J. Casanova}
\affiliation{Department of Physical Chemistry, University of the Basque Country UPV/EHU, 48080 Bilbao, Spain}
\affiliation{EHU Quantum Center, University of the Basque Country, 48940 Leioa, Spain}
\affiliation{IKERBASQUE, Basque Foundation for Science, Plaza Euskadi 5, 48009 Bilbao, Spain}
\author{P. Acedo}
\affiliation{Department of Electronic Technology, Universidad Carlos III de Madrid, Avda. de la Universidad 30, 28911 Legan\'es, Spain}
\author{B. Naydenov}
\affiliation{ Helmholtz-Zentrum Berlin f\"ur Materialien und Energie GmbH, Hahn-Meitner-Platz 1, 14109 Berlin, Germany}
\author{R. Puebla}
\affiliation{Department of Physics, Universidad Carlos III de Madrid, Avda. de la Universidad 30, 28911 Legan\'es, Spain}
\author{E. Torrontegui}
\email{eriktorrontegui@gmail.com}
\affiliation{Department of Physics, Universidad Carlos III de Madrid, Avda. de la Universidad 30, 28911 Legan\'es, Spain}
\email{eriktorrontegui@gmail.com}

\begin{abstract} We model and experimentally demonstrate the full time-dependent counting statistics of photons emitted by a single nitrogen-vacancy (NV) center in diamond under non-resonant laser excitation and resonant microwave control. A generalization of the quantum jump formalism for the seven  electronic states involved in the fast intrinsic dynamics of an NV center provides a self-contained model that allows for the characterization of its emission and clarifies the relation between the quantum system internal states and the measurable detected photon counts. The model allows the elaboration of detection protocols to optimize the energy and time resources while maximizing the system sensitivity to magnetic-field measurements.
\end{abstract}

\maketitle

\vspace{2pc}
\noindent{\it Keywords}: Nitrogen vacancy centers, Full photon statistics.

\section{Introduction}\label{Introduction}

The nitrogen-vacancy (NV) center in diamond has emerged as a prominent platform in the field of quantum technologies \cite{schirhagl14, rondin2014} due to its formidable optical transitions that enable initialization and spin-state discrimination, in addition to exceptional stability, extended coherence time of up to several milliseconds, even at room temperature, and effective control using both microwave (MW) and optical methods \cite{jelezko2006, Jensen_2013}. Its level structure and coherence time are sensitive to different physical fields, making it a versatile sensor with nano-scale spatial resolution and high sensitivity. NV centers have demonstrated great potential in applications of magnetic field \cite{hong2013nanoscale, bao2023quantum}, temperature \cite{hsiao2016fluorescent}, electric field \cite{dolde2011electric}, and strain \cite{ho2021recent} sensing, as well as in nuclear magnetic resonance \cite{mamin2013nanoscale}. Additionally, they can be used as quantum emitters \cite{sipahigil201, leifgen2014}: when optically excited, they can absorb a photon and transition to an excited state, and then relax back to the ground state by the emission of a single photon. The fluorescence emission properties depend on the NV spin state, allowing for a spin-dependent readout~\cite{nizovtsev2001, hopper2018}. Understanding the spin-dependent photon emission is crucial for characterizing the dynamics of the NV center and optimizing its performance in quantum information processing, sensing and communication tasks~\cite{leifgen2014, su2008, rodiek2017}. 

Fluorescence properties of the NV center have been modeled through rate equations for  specific energy levels \cite{Jensen_2013, nizovtsev2001, robledo2011}. This method provides the fluorescence properties based on the populations of the NV electronic states. Different energy models and sets of transition rates have been explored for the NV center in various contexts, including ionization dynamics~\cite{wirtitsch2023exploiting}, hyperfine couplings ~\cite{Cappellaro17, auzinsh2019hyperfine}, level anti-crossing~\cite{ivady2021photoluminescence} and optically detected magnetic resonance (ODMR) experiments~\cite{Jensen_2013}. However, these methods typically either rely on a steady-state approximation of the NV dynamics or simply focus on state populations via rate equations. In either case, these approaches do not directly allow for a precise characterization of the key experimental quantity, namely, the emitted stream of photons~\cite{schmunk2012, temnov2009}.


Given the limitations of current approaches, we make use of a formalism based on quantum jumps \cite{Cook81}. This technique has been previously used to model the emission of different quantum systems such as single molecules \cite{Brown03}, qubit readout of transient charge states \cite{ Anjou2017}, quantum dots \cite{QuantumDot}, electron transport \cite{rudge2019counting}, and NV ionization \cite{Anjou2016maximal}. 
As we experimentally demonstrate, the quantum jump formalism can be adapted to account for the full photon emission statistics (PES) of the NV center within a modelization of the seven electronic states involved in the fast intrinsic dynamics \cite{Manson2006}. The PES are defined as the probabilities $P(n, t)$ of emitting a certain number $n$ of photons at specific time $t$, and serve as a phenomenological framework for establishing the connection between the quantum state of an NV center and the stream of detected photons. The proposed model fully characterizes the emission properties of a single NV, providing physical insight of the NV fluorescence, giving access to different relevant measurable quantities, and allowing for the design of efficient magnetic-field sensing protocols.

This work is structured as follows. The PES model is described in Sect.~\ref{model}, introducing first in Sect.~\ref{energy_levels} the relevant energy levels, and solving their dynamics in Sect.~\ref{jumps} using generalized multi-level Bloch equations, also referred to as a quantum jump formalism. Section~\ref{experiment} presents the experimental setup for benchmark and continues with a set of applications. Section ~\ref{readout} discusses how this statistics allows for optimized state readout, and in Sect.~\ref{g2} we analyze the properties of the emitter by computing the autocorrelation function from the photon statistics. Additionally, we simulate characterization measurements such as the saturation curve of the NV in Sect.~\ref{satcurve}, and optimize the sensitivity in continuous wave ODMRs, which combine both microwave and laser excitation, in Sect.~\ref{cwODMR}. The main conclusions of this work are summarized in Sect.~\ref{conclusions}.

\section{NV photon counting statistics}\label{model}
\subsection{Energy levels}\label{energy_levels}

\begin{figure}[t!]
  \centering
  \includegraphics[height=6cm]{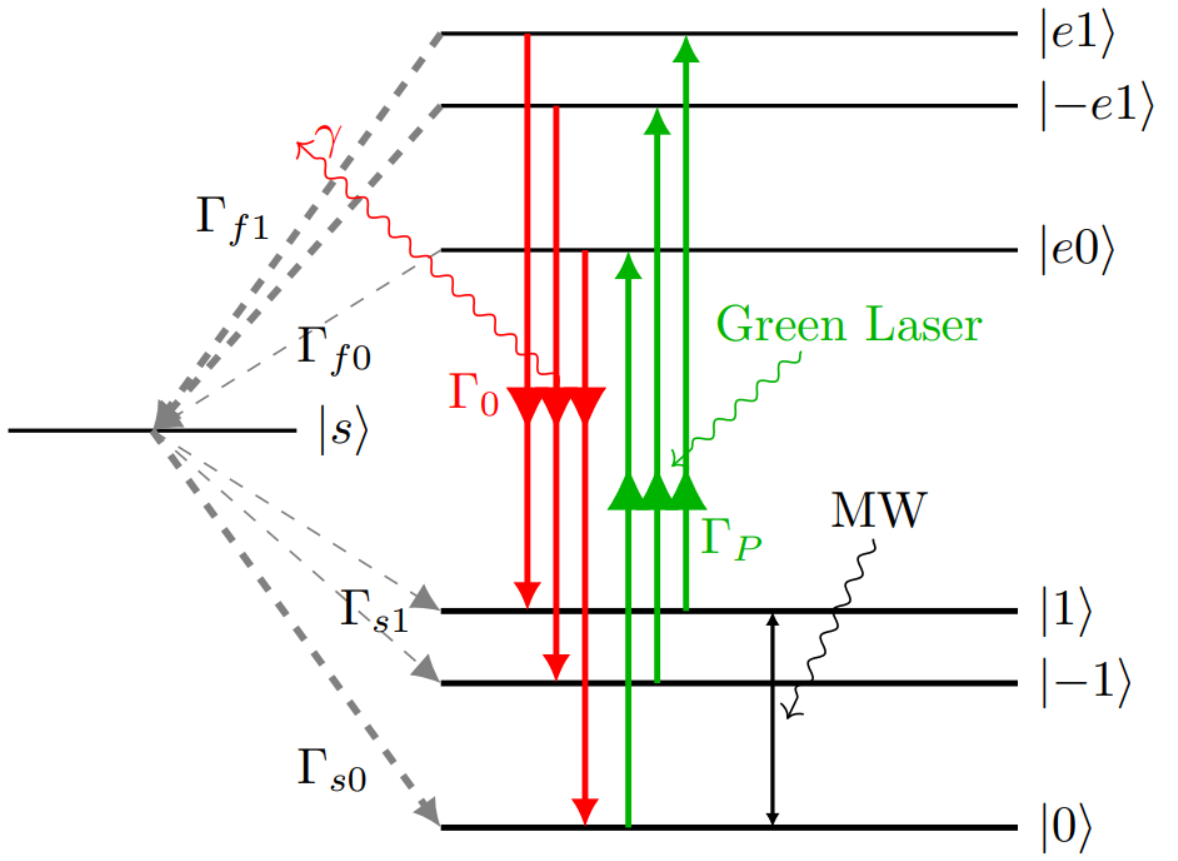}
 \caption{Representation of the levels of an NV center showing the system dynamics. The electronic ground states $\ket{0}$ and $\ket{1}$ encode the qubit and are coupled with a MW field of Rabi frequency $\Omega$ and detuning $\Delta$. Laser light excites the states $\ket{0}$ and $\ket{1}$ with equal rate $\Gamma_P$ to the excited states $\ket{e0}$ and $\ket{e1}$. The excited states decay to the ground state with rate $\Gamma_0$ emitting a photon. The state $\ket{e1}$ also decays to the singlet state $\ket{s}$ with rate $\Gamma_{f1}\gg\Gamma_{f0}$. The state $\ket{s}$ can decay to either ground state $\ket{0}$ or $\ket{1}$ with rates $\Gamma_{s0}$ and $\Gamma_{s1}$, respectively.}
 \label{levels}
\end{figure}

As we are interested in the polarization and readout of a negatively charged NV$^-$ (in the following NV), we consider the fast intrinsic dynamics~\cite{Manson2006} responsible for the radiative and non-radiative transitions between the electronic states. The NV relevant states are distributed in two different spin $S=1$ manifolds, the $^3A_2$ electronic ground state with the $\{ \ket{0}, \ket{1}, \ket{-1} \}$ states used to encode a qubit and the $^3E$ excited manifold containing the $\{ \ket{e0}, \ket{e1}, \ket{-e1} \}$ states. 
Laser light is used to couple the ground states to the spin-triplet excited states. When excited, the NV center relaxes either directly through a spin-conserving
transition emitting red fluorescence of $\sim 637-800$ nm wavelength, or across
a non spin-conserving channel through a meta-stable spin-singlet state $\ket{s} $ responsible for the initialization and readout of the NV. The energy levels
together with the allowed transitions between them are depicted in Fig. \ref{levels}.

The coherent dynamics of the electronic ground state under MW resonant control and a static magnetic field $B_z$ aligned with the NV axis is described by the Hamiltonian
\begin{equation}
\label{Hs}
   \hat H = D\, \hat S_z^2 - \gamma_e\, B_z\, \hat S_z\ + \Omega \hat S_x \cos (\omega t),
\end{equation}
where we have neglected perpendicular terms and hyper-fine interactions. The NV operators $\hat S_{x,y,z}$ correspond to spin-1 matrices. The first term is the intrinsic zero-field splitting with $D=(2\pi) 2.87\,$ GHz. The second is the Zeeman splitting of the $\ket{m_s}=\ket{\pm 1}$ states due to the static magnetic field $B_z$, proportional to the electron gyromagnetic ratio $ \gamma_e = (2\pi) 28.024$ $\frac{\text{MHz}}{\text{mT}}$. The last term represents the MW control with a Rabi frequency $\Omega$ proportional to the square root of the microwave power $\Omega\propto\sqrt{P_{MW}}$. A proper selection of the driving frequency $\omega$ enables the qubit encoding either in the $\ket{0}\leftrightarrow\ket{1}$ or $\ket{0}\leftrightarrow\ket{-1}$ states with transition frequency $\omega_0=D\pm\gamma_eB_z$. This is done by setting the frequency of the MW source close to the transition $\omega=\omega_0+\Delta$, with $\Delta$ a small detuning. Under this condition and moving to an interaction picture with respect to $\hat H_0 = D \hat S_z^2-\gamma_e B_z \hat S_z$, we can neglect transitions to the remaining spin level, and consider our system as a spin-$\frac{1}{2}$ in which the Hamiltonian takes the form
\begin{equation}
  \hat H_2 = \Delta \hat \sigma_z + \Omega \hat \sigma_x,
\end{equation}
with $\hat \sigma_i$ the Pauli matrices. Without loss of generality, let us assume that we select the resonance $\omega = D+\gamma_e B_z$ such the $\ket{-1}$ is left out of the spin dynamics.

%

On the other hand, the non-unitary dynamics of the system, described by the density matrix $\varrho(t)$, follows the Lindblad master equation 
\begin{equation}
  \dot \varrho= -i [\hat H_2, \varrho] + L[\varrho] = \mathcal{L}[\varrho],
  \label{eqn:lioville}
\end{equation}
with 
\begin{equation}
   L[\varrho] = \sum_{k} \Gamma_k(\hat A_{k} \varrho \hat A_{k}^\dagger - \frac{1}{2}\{ \hat A_{k}^\dagger \hat A_{k}, \varrho \}),
\end{equation}
where $\hat A_{k}$ is a jump operator that accounts for the dissipative part of the dynamics with rate $\Gamma_k$ \cite{Cappellaro17}.

Using this master equation, it is possible to extend beyond a basic rate equation model and comprehensively consider the impact of transverse fields and coherent spin polarization exchange. Laser light excites the electronic ground states to the excited states with a rate $\Gamma_{0}^{eo} = \Gamma_{1}^{e1} = \Gamma_{-1}^{-e1} = \Gamma_P$ through a spin conserving transition, where their associated jump operators $\hat{A}_k$ are of the form $\ket{e0}\bra{0}$, $\ket{e1}\bra{1}$ and $\ket{-e1}\bra{-1}$, respectively. The excitation rate $\Gamma_P$ is modeled assuming a linear dependency with the laser power $P_L$, $\Gamma_P = c_L P_L$ \cite{Jensen_2013}. The parameter  $c_L \simeq 0.1$ MHz/$\mu$W, is estimated for our current setup by fitting the PES model to the experimental data through the saturation curve \ref{satcurve}. The electronic excited states can spontaneously decay with rates $\Gamma_{e0}^{0} = \Gamma_{e1}^{1} = \Gamma_{-e1}^{-1} = \Gamma_0$ to the ground states. 

The operators $\hat A_k$ do not only represent incoherent jumps between different states, but can also account for other non unitary processes such as those producing  longitudinal and transverse relaxation of the ground state, with rates $\gamma_1 = 1/T_1$ and $\gamma_2 = 1/T_2$. Without external excitation, the populations of the ground states $\ket{0}$ and $\ket{\pm 1}$ are assumed to decay towards equal populations at a rate $\gamma_1$ through the operators $ \hat A_+ = \hat S^+$ and $ \hat A_-=\hat S^-$ with rate $\Gamma_{\pm}=\sqrt{\gamma_1/2}=\sqrt{1/(2T_1)}$. The transverse relaxation enters via a jump operator $\hat{A}_z=\hat{S}_z$, whose associated rate is given by $\Gamma_z=\sqrt{\gamma_2}=\sqrt{1/T_2}$. 

Optical readout is possible because the $\ket{e1}$ and $\ket{-e1}$ states may also decay to the metastable singlet state $\ket{s}$ with rates $\Gamma_{e1}^{s} = \Gamma_{-e1}^{s} = \Gamma_{f1}$ much faster than the decay from the $\ket{e0}$ to the metastable state $\Gamma_{e0}^{s} = \Gamma_{f0}$. Thus, the $\ket{0}$ state is brighter than the $\ket{\pm 1}$ states when excited with the green source, as more population relaxes back to the ground state through radiative transitions in that case. Furthermore, optical polarization occurs because the state $\ket{s}$ decays with different rates $\Gamma_s^0 = \Gamma_{s0}$ and $\Gamma_{s}^1 = \Gamma_{s}^{-1} = \Gamma_{s1}$ to the ground states $\ket{0}$ and $\ket{\pm1}$. Typical values of the decay rates are $\Gamma_0 \simeq 63$ MHz, $\Gamma_{f0} \simeq 12 $ MHz, $\Gamma_{f1}\simeq 80 $ MHz, $\Gamma_{s1} \simeq 2.4 $ MHz and  $\Gamma_{s0} \simeq 3.3 $ MHz \cite{Cappellaro17}. Similar values can be found in \cite{Jensen_2013, nizovtsev2001, wirtitsch2023exploiting, Robledo_2011}.

%
\subsection{Quantum jumps}\label{jumps}



The master equation given by Eq.~\eqref{eqn:lioville} provides information about the time evolution of the system density matrix, but it does not directly give us information about the number of photons that are emitted during the evolution. To address this, the density matrix $\varrho(t)$ is generalized to include information about the number of spontaneous photon emission events (or jumps) that have occurred prior to time $t$~\cite{Cook81}. This generalization allows us to resolve the full density matrix into individual components $\rho(n, t)$, representing a quantum trajectory with $n$ photons being spontaneously emitted by the NV center during the time interval $[0, t]$,
\begin{equation}
  \varrho(t) = \sum_n  \rho(n, t).
\end{equation}
The equations of motion for these generalized matrix elements follow from Eq.~\eqref{eqn:lioville} via an exact operator expansion for $\rho$ in Liouville space \cite{QuantumDot},
\begin{equation}\label{eq:rhont}
  \dot{\rho}(n, t) = \mathcal{L}_0[\rho(n, t)] + \mathcal{L}_1[\rho(n-1, t)],
\end{equation}
where we split the Liouville operator from Eq.~\eqref{eqn:lioville} $\mathcal{L}[\varrho] = \mathcal{L}_0[\varrho] + \mathcal{L}_1[\varrho] $ in such a way that we have singled out the jump operator associated to the radiative spontaneous decays $\mathcal{L}_1$. This necessarily advances the cumulative photon count by one $\rho(n-1,t) \rightarrow \rho(n,t)$,  with rate $\Gamma_0$, of the excited states $\{ \ket{e0}, \ket{e1}, \ket{-e1} \}$ into the ground states $\{ \ket{0}, \ket{1}, \ket{-1} \}$
\begin{equation}
  \mathcal{L}_1[\varrho] = \Gamma_0 \sum_{i\in \{0, 1, -1\}} (\ketbra{i}{ei} \varrho \ketbra{ei}{i} - \frac{1}{2}\{\ketbra{ei}, \varrho \}),
\end{equation}
and $\mathcal{L}_0[\varrho] = \mathcal{L}[\varrho] - \mathcal{L}_1[\varrho]$.

We numerically solve the set of coupled differential equations (see Appendix \ref{appA}) that dictate the dynamics described by Eq.~\eqref{eq:rhont} under time-dependent laser intensity, MW power and detuning by iterating up to a specific cutoff of emitted photons. Recall that  for smaller and simpler systems the problem can be solved analytically by introducing generating functions \cite{Cook81, Brown03, Anjou2017, QuantumDot}, this is no longer the case for general time-dependent laser and/or MW controls. 

The PES of the system is obtained by tracing $\rho(n, t)$,
\begin{equation}
\label{Pn}
  P(n, t) = \Tr{\rho(n, t)}.
\end{equation}
%

\begin{figure}[t!]%
  \centering
\includegraphics[width= 0.96\linewidth]{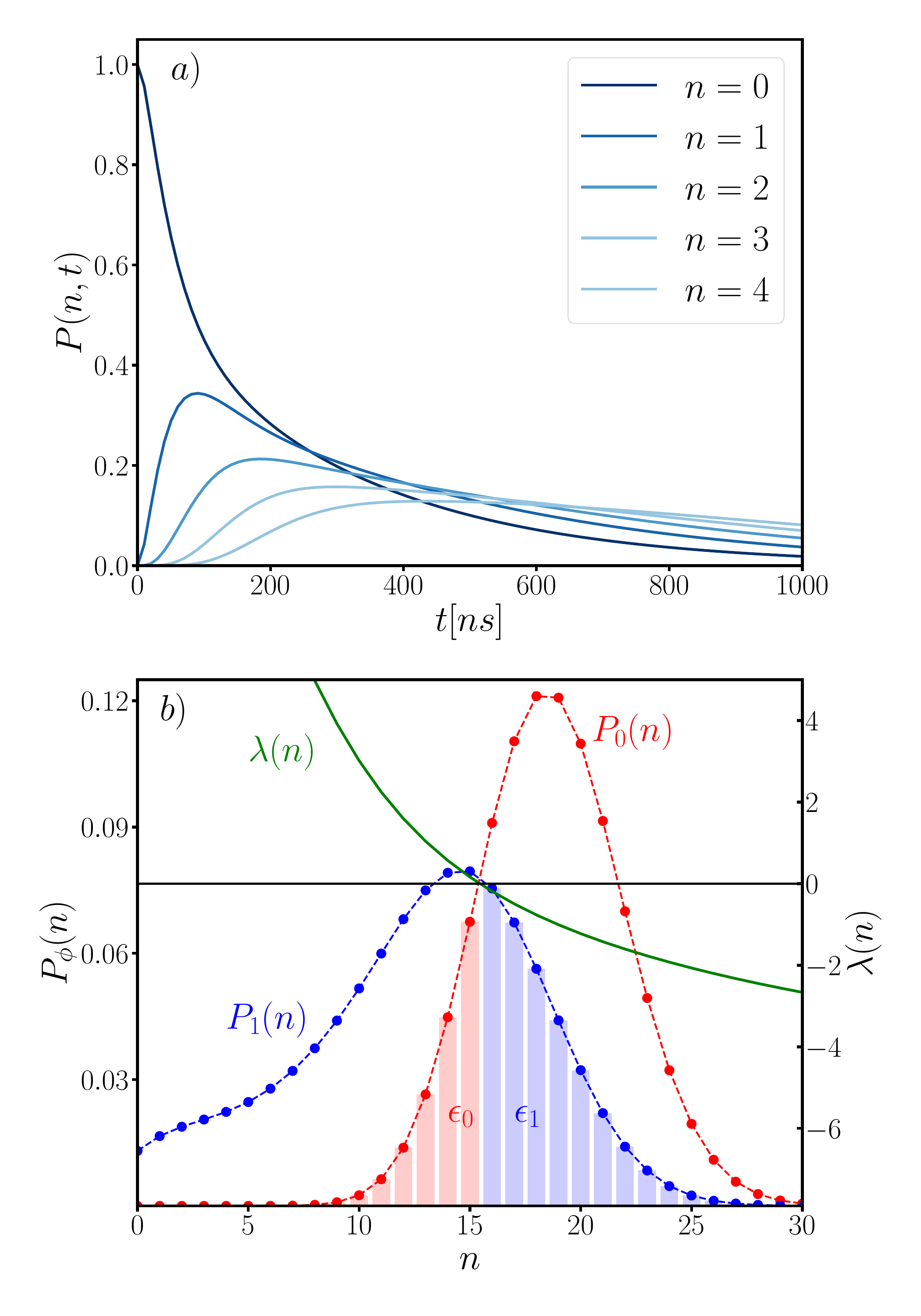}%
   \caption{(a) Emission probability of different number of photons $n$ as a function of time when exciting the NV center initially in a thermal state with a laser source of $\Gamma_P = 10$ MHZ. (b) After 700 ns of laser irradiation, we plot the probability distributions of measuring $n$ photons for the initial state $\ket{0}$ in red ($P_0(n)$) and $\ket{1}$ in blue ($P_1(n)$). The error $\epsilon = (\epsilon_0 + \epsilon_1)/2$ is represented by the area of the overlap between both functions. We plot in green the log-likelihood ratio $\lambda$ given by Eq.~\eqref{eqn:log_ratio}, which can be understood as the value of how confidently we can assign a photon number $n$ to each state.}%
  \label{fig:Pn}%
\end{figure}

This formalism captures the probability distributions of an NV regardless of its initial configuration, going beyond the steady-state configuration. 
Two examples of $P(n,t)$ are represented in Fig.\ref{fig:Pn}a as a function of time for different $n$ and in Fig.\ref{fig:Pn}b as a function of $n$ at a fixed time for two different initializations of the NV center. 
%
%
From these distributions we compute the expectation value of the number of emitted photons as a function of time, as well as higher moments of the distribution, 
\begin{equation}
\label{Nm}
  \expval{n^m(t)} = \sum_n P(n,t)n^m,
\end{equation}
and also derive the counts per seconds, i.e. the photon intensity $\expval{I}$, that the NV center emits through the time derivative of the average number of photons,
\begin{equation}
  \expval{I(t)} = \expval{\dot n (t)}.
  \label{eqn:int_n}
\end{equation}

Employing these  tools, we performed numerical simulations to compute the PES and used them to analyze different physical scenarios of interest. Overall, the model provides a comprehensive framework for understanding the photon emission dynamics of NV centers, and can be used to make predictions under a range of experimental conditions.
\section{Applications}\label{measurements}
\subsection{Experimental methods}\label{experiment}

The experimental measurements to benchmark the model presented in Sect.~\ref{model} were performed on an NV center from a diamond layer grown by chemical vapour deposition on a high pressure high temperature substrate. Optical characterization of NV centers was performed using a custom confocal microscope with  continuous wave excitation for optically detected magnetic resonance (ODMR) experiments. The confocal setup used to interrogate single NV centers is similar to the one used in \cite{Volkova22}. For NV optical excitation we use a diode laser 
with a wavelength of 515 nm. The NV center fluorescence was collected using a $100\times$ oil objective 
with a $1.4$ numerical aperture. Emitted photons were detected by a SPAD 
and counted with a FAST ComTec MCS8. The  
Qudi software~\cite{binder2017qudi} suite was used to control the experiments and for data collection. The static magnetic field was generated with a permanent magnet. The microwave source used for ODMR experiments was TTi TGR600 with the amplifier Mini-Circuits ZHL-16W-43-S+. The AWG used to create the pulses for the Rabi experiment was Tektronix AWG7122C. Both MW signals were delivered to the sample through a cable antenna. All the experiments were performed employing the same selected NV center except for the Rabi oscillations where a second NV with a a $^{15}$N $20-30$ keV implantation was used.  
\subsection{State readout }\label{readout}
NV-based sensing applications rely on standard control sequences that involve spin initialization, manipulation, and readout. The effectiveness of the spin readout process is crucial to the overall performance of the sensor, which can be evaluated within the PES model \cite{D_Anjou_2021}. PES provides a useful framework for proposing new measurement strategies, improving sensitivity based on Bayesian estimation \cite{zhang2023efficient}, performing efficient single shot readout measurements \cite{neumann2010} or speeding them up through a real-time adaptive decision rule \cite{Anjou2016maximal}, which can substantially decrease the average measurement time without significantly affecting readout fidelity. In the context of NV centers, a readout process usually consist on the discrimination between the $\ket{0}$ and $\ket{1}$ states of the NV center that encode the qubit. These states can be distinguished through the number of emitted photons having different statistical properties depending on the internal levels involved in the transition \cite{hopper2018, steiner2010}. 
%

The single-repetition error rate $\epsilon$ is a commonly used figure of merit for the readout performance in a single repetition \cite{D_Anjou_2021}, defined as the average probability in a binary assignment of the incorrect state to the observed outcome. For an ideal measurement with $\epsilon = 0$ the distributions associated to each state do not overlap \cite{degen2017quantum}, so we can discriminate between the $\ket{0}$ and $\ket{1}$ states in a single readout. However, a readout is subject to uncertainties in the discrimination leading to non-ideal measurements and overlap between the probability distributions $P_0(n,t)$ and $P_1(n,t)$ that hinder the $\{\ket{0},\ket{1}\}$ state discrimination, see Fig. \ref{fig:Pn}b. The assignment rule that minimizes $\epsilon$ is obtained by calculating the log-likelihood ratio
\begin{equation}
  \lambda(n, t) = \ln{\frac{P_{0}(n, t)}{P_{1}(n, t)}}. \label{eqn:log_ratio}
\end{equation}
When $\lambda(n, t)$ is larger (smaller) than 0, the state $\ket{1}$ ($\ket{0}$) is assigned, see Fig. \ref{fig:Pn}b. The log-likelihood ratio can be interpreted as the level of confidence of the observer in the state assignment given the collected number of photons $n$. The average single-repetition error rate is $\bar\epsilon = \frac{1}{2}(\epsilon_0 + \epsilon_1)$, where
\begin{equation}
  \epsilon_0(t) = \sum_{n\leq n_c} P_0(n, t), \hspace{1cm}  \epsilon_1(t) = \sum_{n > n_c} P_1(n, t),
\end{equation}
are the error rates conditioned on preparation of $\ket{0}$ and $\ket{1}$ with $n_c$ defined such as $\lambda(n_c, t) = 0$. These error rates $\epsilon_{0}$ and $\epsilon_{1}$ are represented by color bars in Fig. \ref{fig:Pn}b. 

Although a single readout cannot determine with certainty whether the NV center is in the $\ket{0}$ or $\ket{1}$ state due to measurement uncertainty, repeated readouts “average out” the noise, mitigating readout errors. By performing $N$ repetitions of the readout process provides a set of photon counts outcomes $\{ n_1, n_2, ... n_N \}$ 
with an associated cumulative error rate $e_N$, that decreases asymptotically as the number of measurements $N$ grows
\cite{D_Anjou_2021}
\begin{equation}
  \ln e_N(t) \rightarrow - C(t)N  \text{ as }  N \rightarrow \infty.
\end{equation}
with
\begin{equation}
  C(t) = - \min_{0 \leq s \leq 1} \ln{[\sum_n P_0(n, t)^sP_1(n, t)^{1-s}]}.
  \label{Chernoff}
\end{equation}
The quantity $C$ is known as the Chernoff information \cite{chernoff1952}, a symmetric distance measure between the distributions $P_0(n,t)$ and $P_1(n,t)$ that can be interpreted as a rate of information gain per repetition. 
Readout outcome distributions $P_0(n, t)$ and $P_1(n, t)$ with the same single repetition error rates $\epsilon$ do not necessarily have the same Chernoff information.

\begin{figure}%
  \centering
\includegraphics[width= 0.96\linewidth]{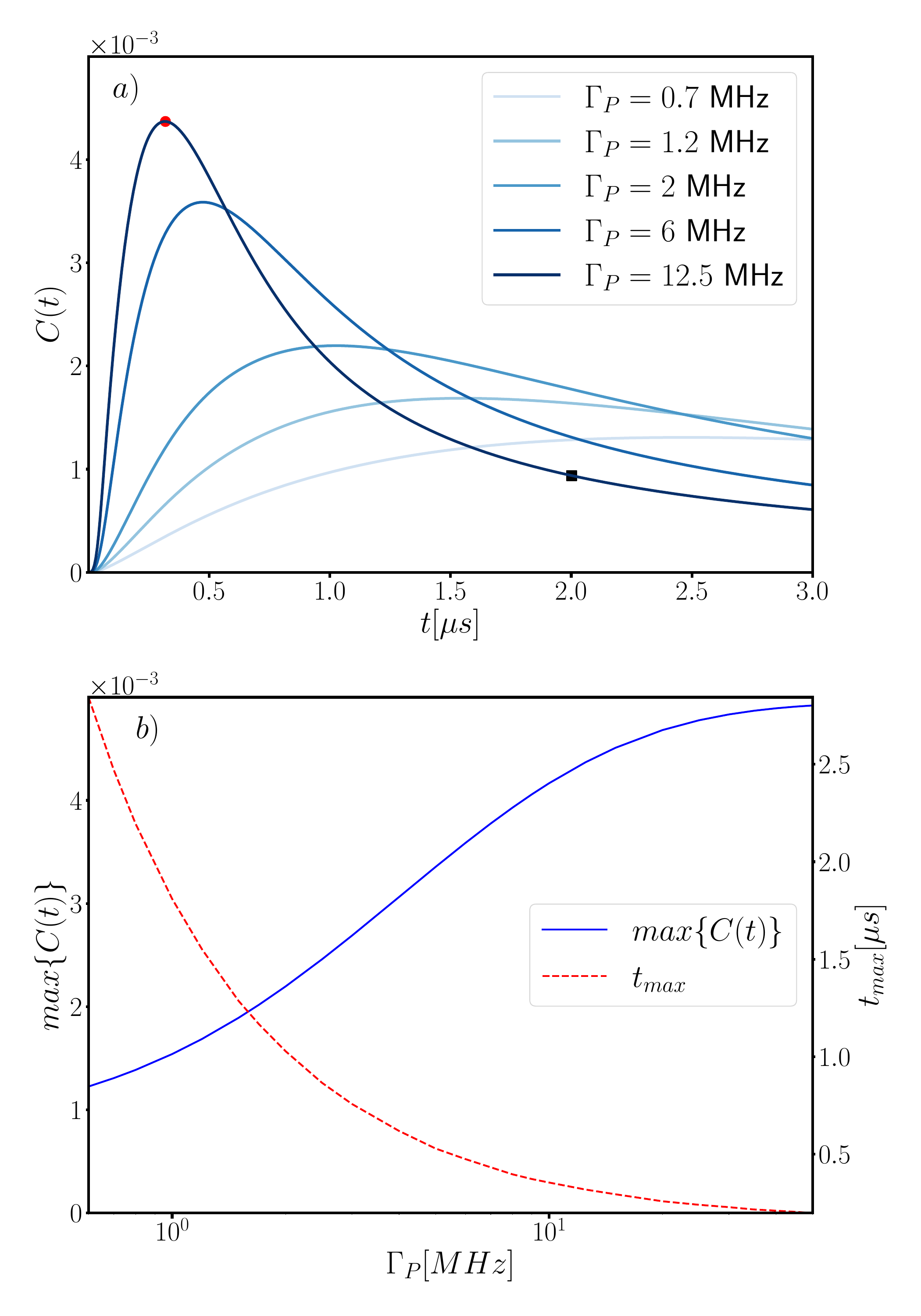} %
  \caption{(a) Calculated Chernoff information for different laser powers as a function of time. The read circle and black square correspond to the pulse power and length at which we simulate two different Rabi oscillations in Fig. \ref{fig:rabinoise}b. (b) The maximum value achieved by the Chernoff information $max \{C(t)\}$, and the time at which this value is achieved $t_{max}$, as a function of the laser intensity. This allows to investigate how fast and well we are able to discriminate the NV center state and the required power. The maximum value and time saturates as the power increases.}%
  \label{fig:Chernoff_max}%
\end{figure}

The calculated Chernoff information is a useful tool for analyzing the statistical properties of the fluorescence. It can be used to estimate the minimum number of photons that need to be measured in order to discriminate between two different quantum states of a light source with a certain level of confidence \cite{D_Anjou_2021, audenaert2007}, thereby providing us with the sensitivity limits of a quantum system. Thus, the Chernoff information provides useful insight into the optimal experimental parameters for maximizing state discrimination. 

\begin{figure}[t]%
  \centering
  \includegraphics[width= 0.96\linewidth]{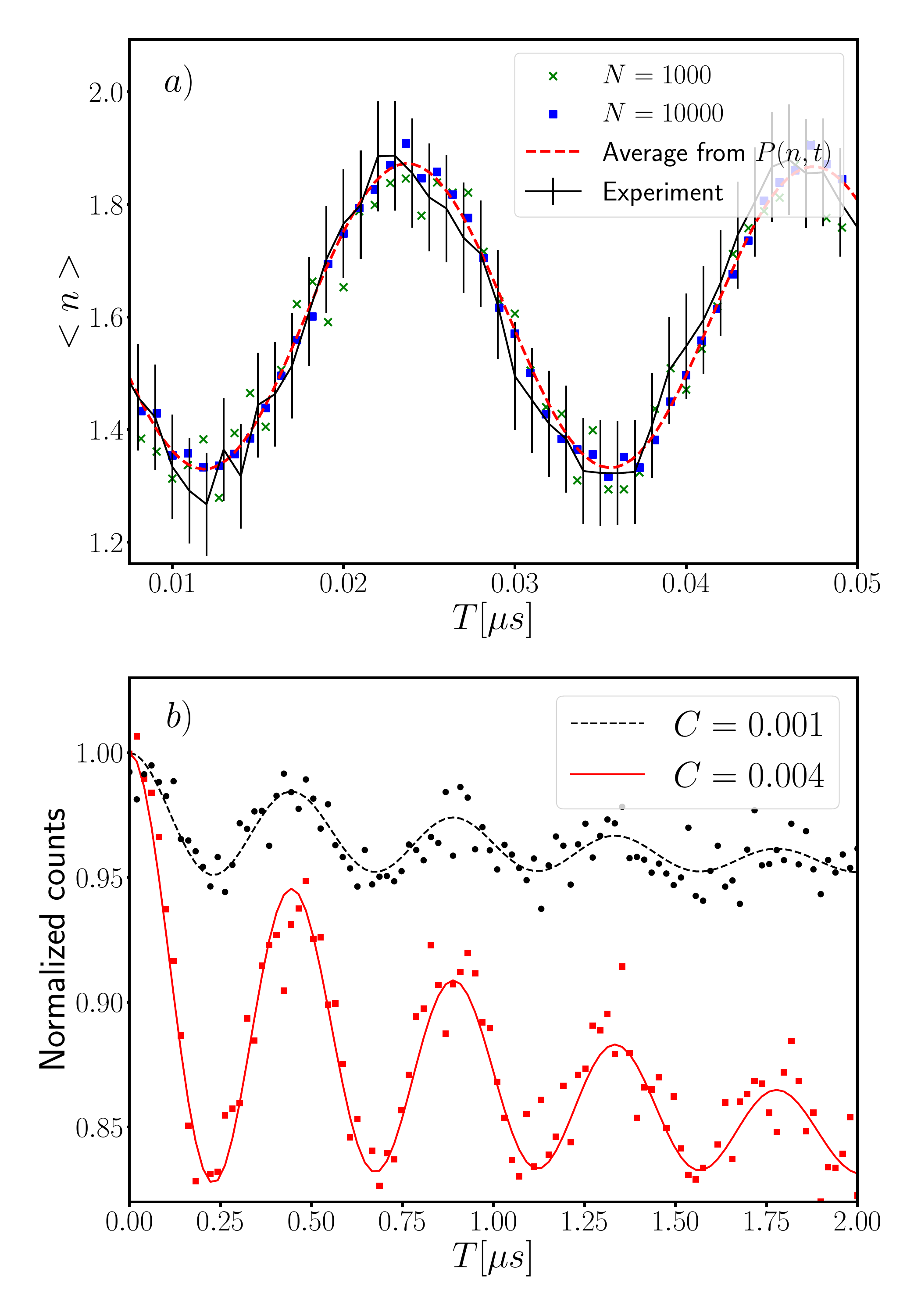}
  \caption{(a) Measured Rabi oscillations for initialization and readout pulses of $2$ $\mu$s and a MW field with $\Omega = 270$ MHz, $\Delta = 0$ MHz and $\Gamma_{P} = 20$ MHz. The dashed line represents the mean photon number $\langle n \rangle$ computed by Eq. \eqref{Nm} from the theoretical $P(n,t)$, symbols correspond to sampling $N$ photon counts, while the black-solid line is the measured Rabi oscillations using the procedure and setup described in Sect.~\ref{experiment}.
  (b) Simulated Rabi oscillations  for different Chernoff informations with $\Omega = \Delta = 10$ MHz and readout excitation rate $\Gamma_P = 12.5$ MHz for two different time pulses. In red $t = 300$ ns corresponding to the Chernoff maximum $C = 4\cdot 10^{-3}$ and in black $t = 2$ $\mu s$ for $C = 1\cdot 10^{-3}$. We plot the normalized counts deduced from the same emission probability $P(n,t)$ considering $N = 3000$ measurements (points) and its average (lines).} 
  \label{fig:rabinoise}%
\end{figure}

\begin{figure*}[t!]%
  \centering
  \includegraphics[width= 0.96\linewidth]{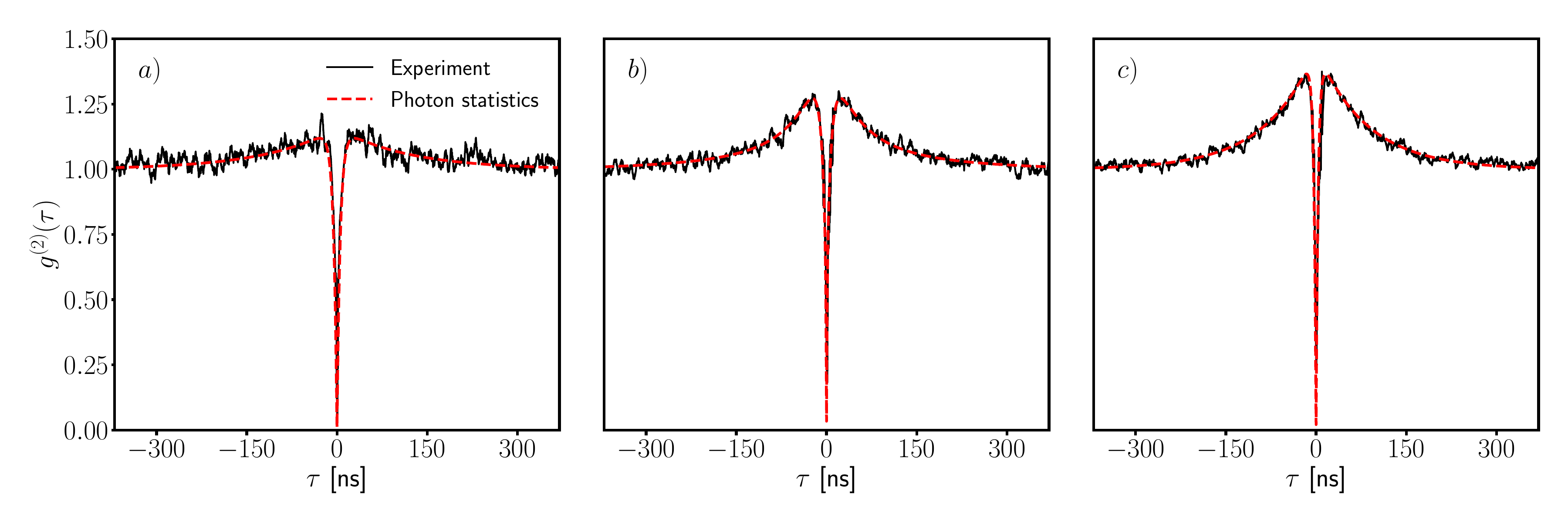}
  \caption{Measured auto-correlation functions (black solid line) and the photon statistics model predictions (red dashed line) according to \eqref{eqn:g2} for different excitation rates: (a) $\Gamma_P = 5$ MHz, (b) $\Gamma_P = 15$ MHz and (c) $\Gamma_P = 20$ MHz }%
  \label{fig:g2}%
\end{figure*}

In Fig. \ref{fig:Chernoff_max}a, the Chernoff information is plotted for different laser intensities as a function of time taking a constant laser irradiation. For a specific laser intensity, the figure shows a maximum, indicating the optimal duration of the laser pulse for reading out the ground state of the NV center. In Fig. \ref{fig:Chernoff_max}b, the maximum value of the peak and the time $t_{max}$ at which it occurs are plotted for different values of $\Gamma_P$. As the laser intensity is increased, the peak becomes higher and occurs faster, resulting in better and faster state discrimination. However, it also increases the background noise introduced in the measurement. This will be discussed in Sect. \ref{satcurve}. Both the Chernoff peak and $t_{max}$ saturate for high laser intensities, meaning that increasing the laser power beyond this point will not result in better measurement outcomes.

To illustrate the relevance of the Chernoff information we have used the PES to simulate the readout of Rabi oscillations. A MW pulse takes the NV to a superposition state, that after a readout laser pulse emits fluorescence with a $P(n,t)$ distribution for a given irradiation time $t$. A photon measurement $n$ can be simulated by sampling a point from this distribution. Thus, $N$ measurements consist on sampling a set of readouts outcomes $\{ n_1, n_2, ..., n_N\}$. On the one hand, as the number of repetitions increases (see symbols of Fig. \ref{fig:rabinoise}a) the distribution of the sampled photons converges to
the average of the distributions (red-dashed line), since the error of repeated measurements averages
out the uncertainties. Note, how the model distribution perfectly cast the experimental data (black-solid line). On the other hand, working conditions for a larger Chernoff parameter (see Fig. \ref{fig:Chernoff_max}a) favors the discrimination of the $\ket{0}$ and $\ket{1}$ states, enhancing the Rabi oscillations amplitude, cf. Fig. \ref{fig:rabinoise}b. The increased amplitude of the Rabi oscillations resulting from larger Chernoff value leads to improved discrimination of the state of the NV center, higher readout fidelity, and ultimately, better experimental outcomes. As seen, the photon statistic model allows to predict and improve measurement sequences by optimizing relevant experimental parameters. 

\subsection{Autocorrelation measurements}\label{g2}
Autocorrelation measurements are a valuable technique for studying the temporal dynamics of fluorescence signals and the statistical behavior of photons in a light source \cite{fishman2021photon, buller2009}. By measuring the correlation between the intensity of the signal at different delay times,
it is possible to extract information about the decay rates of the states of an NV center \cite{Doherty13}, categorize an emitter as a single photon source \cite{morrow2022noisy}, investigate NV centers embedded in different physical structures \cite{Volkova22, andersen2017ultrabright, berthel2015photophysics, albrecht2014narrow} and characterize the interactions with other quantum systems \cite{sipahigil201, englund2010,siampour2017, faraon2012coupling}.  
The second-order temporal coherence $g^{(2)}(\tau)$ measures the photon autocorrelation, that, in terms of the intensity $I(t)$, reads
\begin{equation}
\label{g2I}
  g^{(2)}(\tau) = \frac{\expval{I(t) I(t + \tau)}}{\expval{I(t)}^2}.
\end{equation}

Alternatively, it can be also computed using the delay function $D_1(t)$ describing the distribution of delay times before the next emission occurs. This function is calculated from the survival probability $P(0, t)$ as \cite{Plenio_1998}
\begin{equation}
    D_1(t) = - \dot P(0,t).
\end{equation}
The conditional probability of an NV center emitting any photon between time $\tau$ and $\tau + d\tau$, after having emitted a photon at time $\tau_{0}$ = 0 is $D(\tau) d\tau$. The photon emitted at time $\tau$ can be the first to be emitted after that at time $\tau_{0} = 0$ or the next after any other $\tau'$ ($0 < \tau' < \tau)$, so that
\begin{equation}
  D(\tau) = D_1(\tau) + \int_0^\tau d\tau' D(\tau')D_1(\tau-\tau').
\end{equation}
The function $g^{(2)}(\tau)$ is the normalized correlation function for a photon detection at $\tau_{0} = 0$ followed by the detection of any photon (not necessarily the next) at time $\tau$. It follows that
\begin{equation}
  g^{(2)}(\tau) = \frac{D(\tau)}{\lim_{\tau\to\infty}D(\tau)}.
  \label{eqn:g2}
\end{equation}
In Fig.~\ref{fig:g2} we plot the background noise corrected $g^2(\tau)$ \cite{vorobyov2017superconducting} from a single NV center in a Hanbury-Brown-Twiss correlation experiment \cite{faraon2012coupling}. The results predicted from the photon statistics model \eqref{eqn:g2} perfectly fit the experimental data.

Alternatively, the Mandel $Q$ parameter \cite{Brown03}

\begin{equation}
  Q(t) = \frac{\expval{n^2(t)} - \expval{n(t)}^2}{\expval{n(t)}} - 1,
  \label{eq:mandel}
\end{equation}
is another measurement for the correlation of emitted photons that relies in the variance of the photon number in the detected fluorescence. When $Q = 0$, the counting statistics $P(n, t)$ follow Poisson distributions with no correlation between emitted photons. On the other hand, when $Q>0$, the statistics is super-Poissonian, indicating photon bunching, where photons tend to be emitted together. Conversely, sub-Poissonian statistics ($Q < 0$) indicate photon anti-bunching, where photons tend to be emitted at well-separated times.

\begin{figure}[t!]%
  \centering
  \includegraphics[width= 0.96\linewidth]{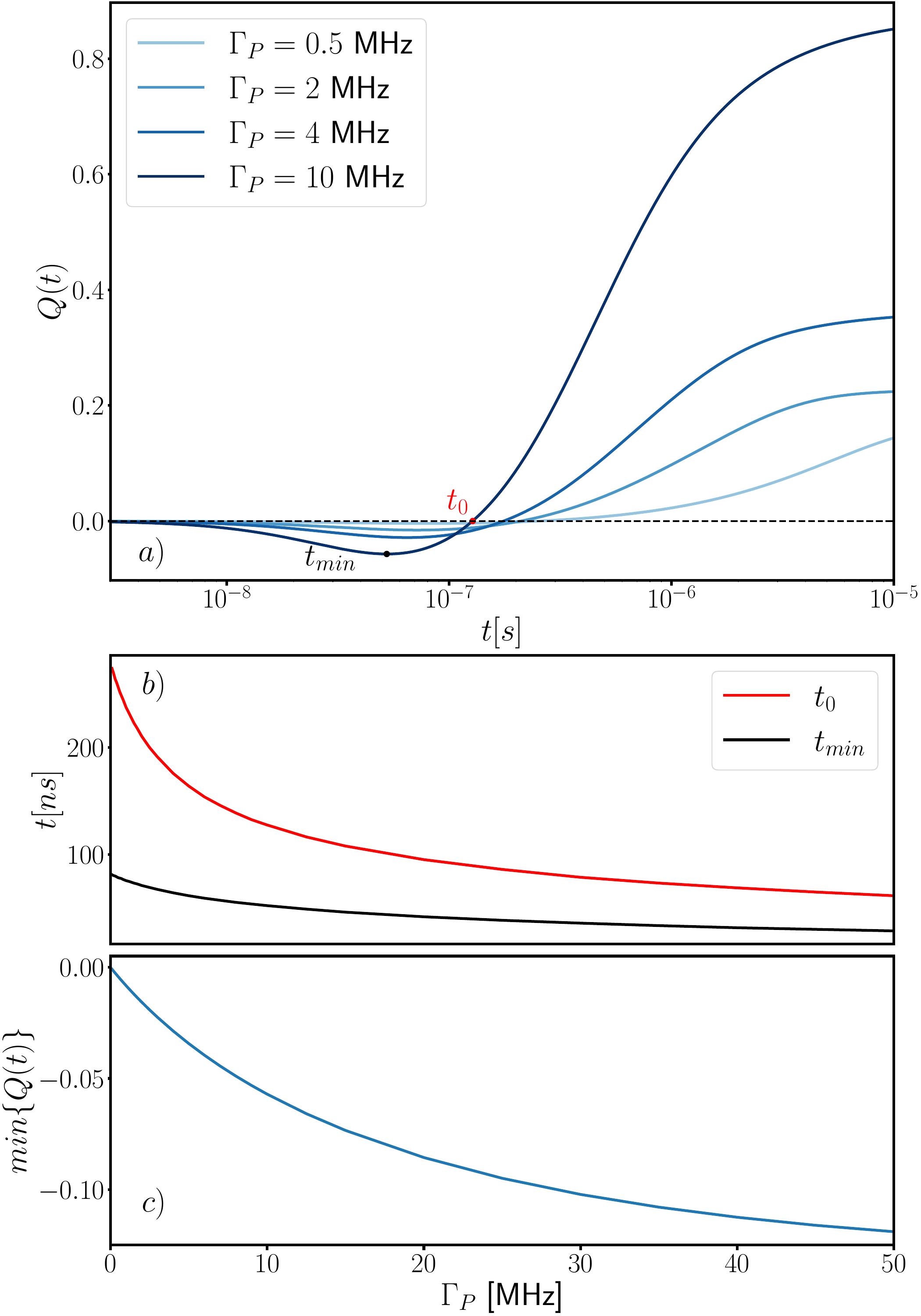}
  \caption{(a) Mandel $Q$ parameter \eqref{eq:mandel} as a function of time for a thermal initialization of the NV center. The characteristic times $t_{min}$ and $t_0$ correspond to a Mandel paramater value $Q(t_{min})=\min\{Q(t)\}$ and $Q(t_0)=0$ of the photon emission, respectively. (b) Both $t_{min}$ and $t_0$, and (c) $Q(t_{min})$ as a function of the laser power.} \label{fig:mandelQ}%
\end{figure}

Using PES, in Fig. \ref{fig:mandelQ}a we calculate $Q(t)$ at different power excitation, observing different bunching behaviours as a function of time for a thermal initialization of the NV center. The photon emission begins with some degree of anti-bunching because the excited states are not yet populated. At time $t_{min}$, such that $Q(t_{min}) = \min \{ Q(t) \}$, the emitted photons are maximally anticorrelated, showing a non-classical anti-bunched character. Then, the statistics starts to bunch at time $t_0$, until it saturates at long times and the emitted distribution of photons becomes thermal. As presented in Figs. \ref{fig:mandelQ}b and \ref{fig:mandelQ}c, the model predicts a higher degree of anti-bunching for increasing power, albeit the time interval during which anti-bunching takes place is reduced. 

This information is helpful to investigate the capabilities of NV centers as photon emitters since it has been found that NV centers exhibit both bunching and antibunching behavior depending on the specific excitation conditions \cite{nizovtsev2001, Grangier00}. Bunching and antibunching phenomena have important consequences on the performance of devices that use quantum systems as a source of photons, such as those used in quantum communication. For example, bunching can lead to increased noise in the emitted light \cite{burkard2000, lachs1968}, while antibunching can improve the use of NV centers for single photon generation \cite{beveratos2002, aharonovich2009, schroder2011}.

\subsection{Fluorescence saturation}\label{satcurve}
As the laser power is increased, the NV center absorbs more photons and becomes excited more frequently. Eventually, as the laser power continues to increase, the populations of the excited state and the ground state become approximately equal, resulting in a saturation of the emission rate of the NV center. Using PES to compute the saturation curve for an NV center allows to optimize and compare the performance of an NV-based sensor~\cite{li2017, obydennov2018}, study its quantum efficiency~\cite{radko2016} and its signal-to-noise ratio~\cite{Gupta_2016}.

\begin{figure}[t!]%
  \centering
  \includegraphics[width= 0.96\linewidth]{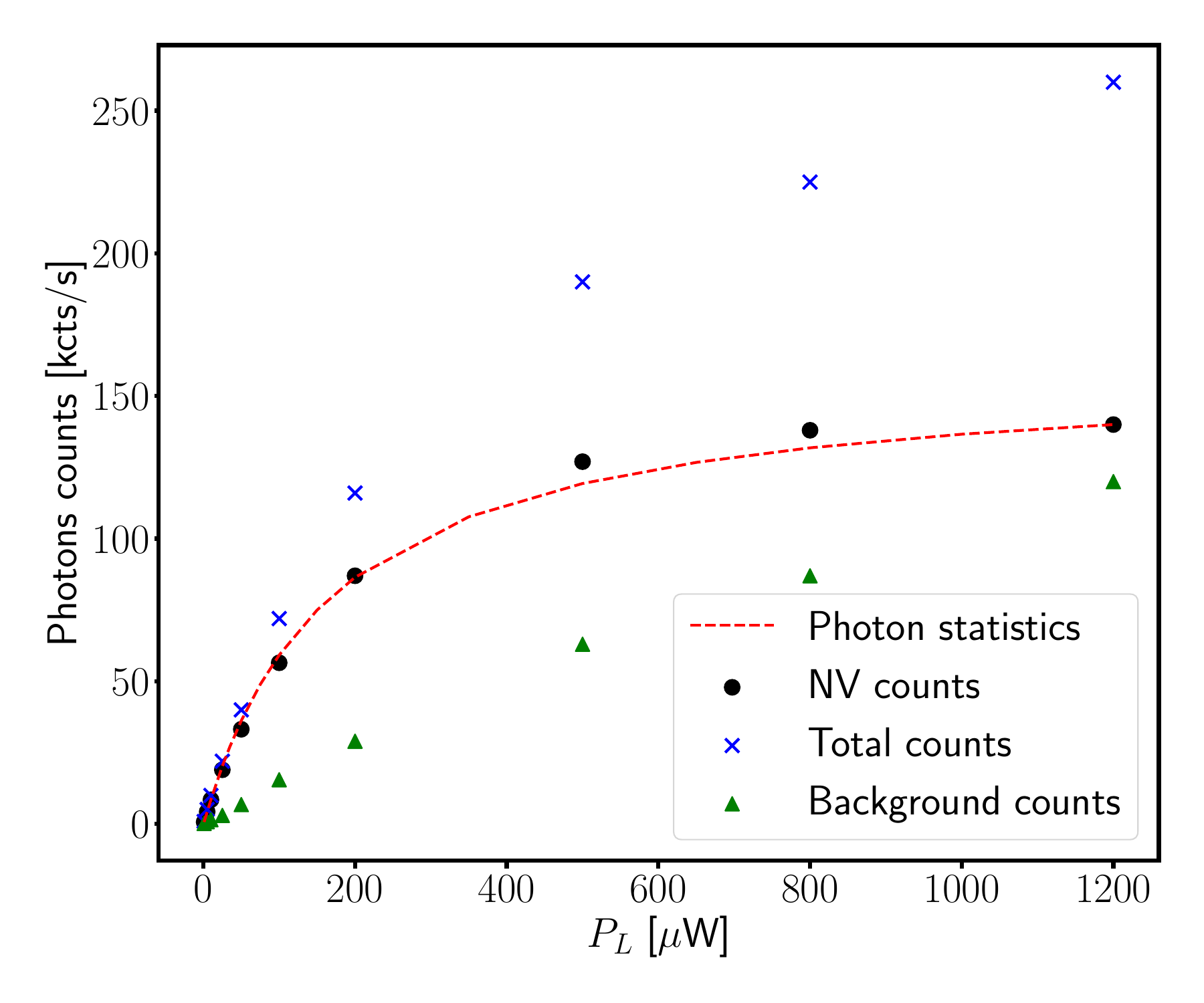}
  \caption{Benchmark of the predicted saturation curve for an NV center as a function of the laser power computed by the photon statistic model (red line) against the detector photon counts (black points). The NV counts were computed by subtracting the background counts (green triangles) to the total measured counts (blue crosses).}%
  \label{fig:satcurve}%
\end{figure}

In Fig. \ref{fig:satcurve} we plot the experimental saturation curve of the selected NV center (black dots), on top of the theoretical curve using the photon statistic model (dashed-red line). Note that the photon statistic model 
matches the NV counts computed by subtracting the background from the total measured counts. We also have taken into account that not all the emitted photons are collected due to the geometry of the experiment and the efficiency of the detector by multiplying the saturation curve by a scaling factor. This scaling factor provides insight into the collection efficiency of the setup and indicates improvement potential. Modeling of such saturation curves may be useful in scenarios where subtracting the background from the total counts can be harder, for instance in experiments with NV ensembles.


\subsection{Continuous Wave Optically Detected Magnetic Resonance (cwODMR)}\label{cwODMR}

The NV centers in diamond have demonstrated remarkable potential as a tool for detecting magnetic fields.  This ability is based on the NV electronic spin level structure and the Zeeman effect, which results in a lifting of degeneracy between the states $\ket{\pm 1}$ when a magnetic field is applied, as discussed in Sect. \ref{model}. The energy splitting between these states is proportional to the applied magnetic field and can be measured via optically detected magnetic resonance (ODMR) experiments.

In an ODMR experiment, a MW field is applied to the NV center while a green laser initializes the NV center to the $\ket{0}$ state. When the MW field is resonant with the transition between the $\ket{0}$ and one of the $\ket{\pm 1}$ states, fluorescence from the NV center drops, as population is transferred from the bright $\ket{0}$ state to the darker states $\ket{\pm 1}$~\cite{Jensen_2013, Doherty13}. This phenomenon enables the detection of magnetic fields by measuring the resonant transition frequency between the $\ket{0}$ and $\ket{\pm 1}$ states.

We perform Continuous Wave ODMR (cwODMR) experiments to benchmark the photon statistic model. These experiments involve the continuous application of both the laser and the MW field to the NV center. While Pulsed ODMR (pODMR) experiments provide greater sensitivity, they require the generation and synchronization of fast MW and laser pulses, which can be demanding in terms of experimental setup. The advantages of cwODMR experiments include their simplicity, ability to measure a wide range of magnetic fields, and capability to detect weak magnetic fields. As a result, the cwODMR method is a valuable technique for a broad range of applications, such as sensing magnetic fields from biological systems \cite{schirhagl14}, imaging magnetic materials \cite{maletinsky2011}, and detecting single-spin dynamics \cite{jacques2008}.

\begin{figure}[t]%
  \centering
\includegraphics[width= 0.96\linewidth]{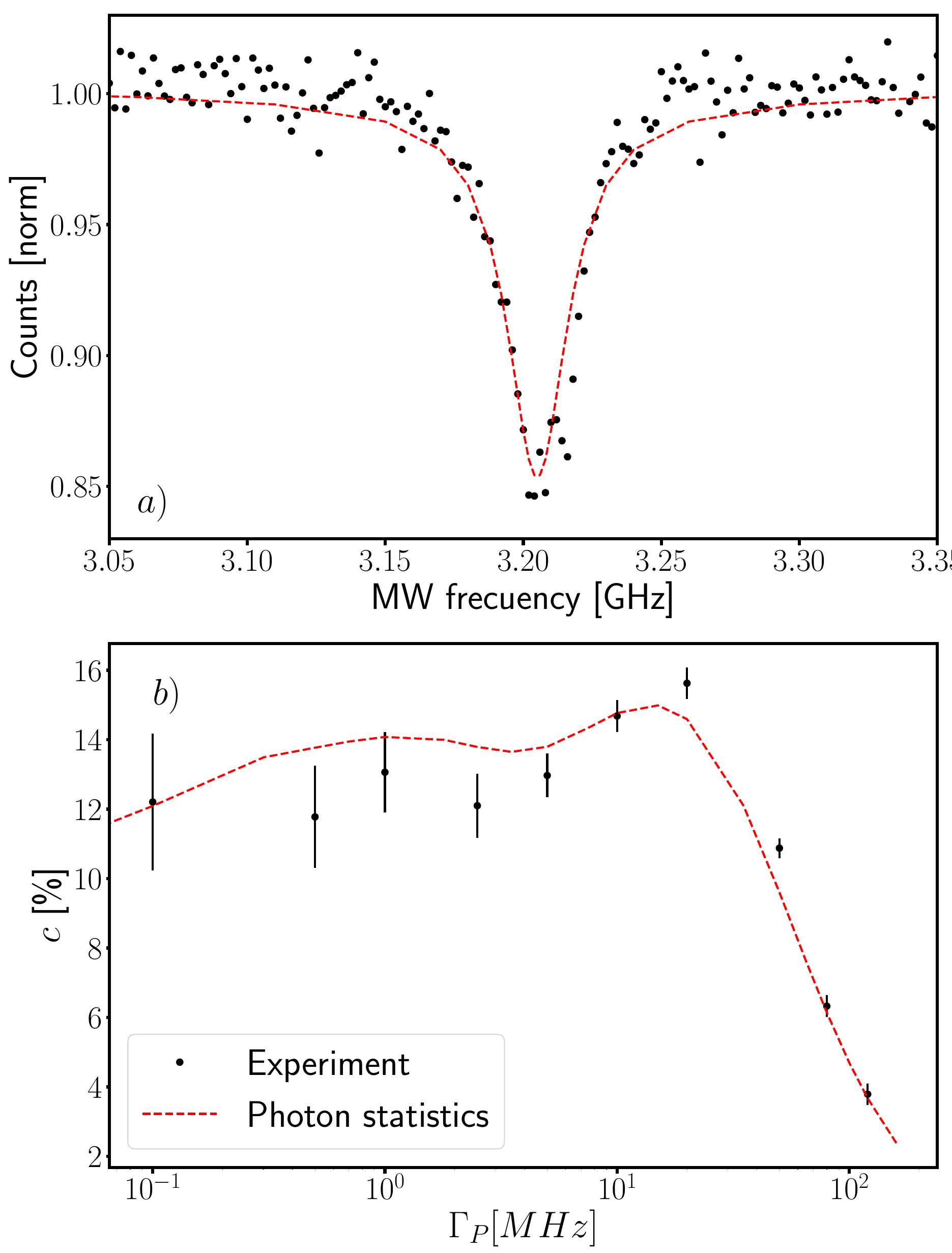} %
  \caption{(a) The black points correspond to an ODMR experiment with a MW power of $\Omega  = 8$ MHz and an excitation rate of $\Gamma_P = 20$ MHz. We obtained a contrast $c = 15\%$. 
  The red line is the result from the photon statistics simulation. (b) Contrast as a function of the laser power at a fixed MW power of $\Omega  = 8$ MHz. The black points correspond to the values from Lorentzian fits to the experimental data and the red line the photon statistics model.}%
  \label{fig:ODMR}%
\end{figure}
The accurate estimation of cwODMR center resonance frequency and precise computation of the corresponding magnetic field requires a high contrast $c$ \cite{Jensen_2013}, quantity that depend on the MW and laser powers, as well as other physical parameters. The photon statistics model predicts the behavior of $c$ allowing the optimization of MW and laser powers that leads to the maximal sensitivity of the NV. Note that the model was successfully used to simulate cwODMR, as demonstrated in Fig.~\ref{fig:ODMR}a. In Figure \ref{fig:ODMR}b, the contrast is plotted for different laser powers, with a fixed MW power of $\Omega  = 8$ MHz. The contrast increases at lower powers and reaches a maximum around $200$ $\mu$W before decreasing. This maximum depends on the applied MW field and occurs due to the NV center being optically excited at a faster rate than the spin transitions. 

\section{Conclusions}\label{conclusions}

The work provides physical insight to the photon emission of an NV allowing its full characterization. The experimental benchmark validates the presented PES model for the fast intrinsic dynamics of the NV center. Within this framework, we were able to relate the qubit state of an NV with the detected stream of photons, and investigate the NV properties in a variety of scenarios, including the state readout  optimization through the Chernoff information and exemplified with the Rabi oscillations, characterizing experimental autocorrelation measurements, and examine the properties of the emitted photons by the bunching and antibunching regimes. Finally, the model allows us to determine the optimal regimes for laser and MW intensities in magnetic sensing experiments. 
Originally proposed to account for the PES of a two-level atom, the method introduced in~\cite{Cook81} can be extended to include the relevant seven-electronic levels of an NV center. This technique can be adapted to benchmark other color centers and quantum systems, as well as to include other phenomena such as ionization \cite{wirtitsch2023exploiting} or two photon absorption \cite{wee2007two}. Additionally, our work can also help to refine current measurements on the NV center energy levels, leading to improved accuracy and precision in future experiments. Furthermore, PES are useful for proposing and evaluating new readout techniques \cite{tian2023bayesian} to improve the NV center sensitivity. Overall, our findings contributes to the understanding of NV centers and have potential implications for quantum sensing and related applications.

\section{Acknowledgements}
We thank J. J. Garc{\'i}a-Ripoll for fruitful discussions and T. Pregnolato and T. Schr\"oder from Ferdinand-Braun-Institute for the $^{15}$N implantation on the diamond sample. This work is supported by Arquimea Research Center and by Horizon Europe, Teaming for Excellence, under grant agreement No 101059999, project QCircle. We acknowledge financial support form the Spanish Government via the projects PID2021-126694NA-C22 and PID2021-126694NB-C21 (MCIU/AEI/FEDER, EU), the German Research Foundation DFG (project numbers 410866378 and 410866565) and  the German Federal Ministry of Education and Research BMBF (Project DIQTOK, number 16KISQ034K), the ELKARTEK project Dispositivos en Tecnolog\'i{a}s Cu\'{a}nticas (KK-2022/00062), the Basque Government grant IT1470-22, by Comunidad de
Madrid-EPUC3M14. H.E. acknowledges the Spanish Ministry of Science,
Innovation and Universities for funding through the FPU
program (FPU20/03409). J. C and E.T. acknowledge the Ram\'on y Cajal (RYC2018-025197-I and RYC2020-030060-I) research fellowship.

\appendix
\section{Generalized Bloch equations}
\label{appA}

The Lindblad equation~\eqref{eq:rhont} can be explicitly expanded into the density matrix projectors $ \rho_{ij}(n,t)\equiv  \rho_{ij}^n$ leading to a set of generalized Bloch equations. Defining the populations  $\rho_{ii}(n,t) = \rho_{i}^n$ these equations read

\begin{align}
        \dot{ \rho}_{0}^n &= -i\frac{\Omega}{2}( \rho_{01}^n -  \rho_{10}^n) - \frac{\gamma_1}{2}( \rho_{0}^n -  \rho_{1}^n) 
        - \frac{\gamma_1}{2}( \rho_{0}^n -  \rho_{-1}^n) \nonumber \\ & - \Gamma_P \rho_{0}^n + \Gamma_0  \rho_{e0}^{n-1} + \Gamma_{s0}  \rho_{s}^n,\nonumber \\
        \dot{ \rho}_{1}^n &= i\frac{\Omega}{2}( \rho_{01}^n -  \rho_{10}^n) + \frac{\gamma_1}{2}( \rho_{0}^n -  \rho_{1}^n) \nonumber \\ & - \Gamma_P \rho_{1}^n + \Gamma_0  \rho_{e1}^{n-1} + \Gamma_{s1} \rho_{s}^n,\nonumber\\
        \dot{ \rho}_{-1}^n &= \frac{\gamma_1}{2}(\rho_{0}^n - \rho_{-1}^n)
        - \Gamma_P\rho_{1}^n + \Gamma_0 \rho_{-e1}^{n-1} + \Gamma_{s1}\rho_{s}^n,\nonumber \\
        \dot{ \rho}_{01}^n &= - (\frac{\gamma_1}{2} + \frac{\gamma_2}{2} + \Gamma_P - i\Delta)\rho_{01}^n + i \frac{\Omega}{2}(\rho_{1}^n - \rho_{0}^n),\nonumber \\
        \dot{\rho}_{10}^n &= - (\frac{\gamma_1}{2} + \frac{\gamma_2}{2} + \Gamma_P + i\Delta)\rho_{10}^n - i \frac{\Omega}{2}(\rho_{1}^n - \rho_{0}^n),\nonumber \\
        \dot{ \rho}_{e0}^n &= \Gamma_P \rho_{0}^n - \Gamma_0 \rho_{e0}^n - \Gamma_{f0} \rho_{e0}^n, \nonumber\\
        \dot{ \rho}_{e1}^n &= \Gamma_P \rho_{1}^n - \Gamma_0 \rho_{e1}^n - \Gamma_{f1} \rho_{e1}^n, \nonumber\\
        \dot{ \rho}_{-e1}^n &= \Gamma_P \rho_{-1}^n - \Gamma_0 \rho_{-e1}^n - \Gamma_{f1} \rho_{-e1}^n,\nonumber \\
        \dot \rho_{s}^n &=  \Gamma_{f1} \rho_{e1}^n + \Gamma_{f1} \rho_{-e1}^n + \Gamma_{f0} \rho_{e0}^n - (\Gamma_{s0} + \Gamma_{s1})\rho_{s}^n .
    \label{eqn:Bloch5n}
\end{align}
From the three first equations, note that the emission of $n$ photons is affected by the $n-1$ photon process, where the state decays from the electronic $^3E$ manifold to the corresponding $^3A_2$ electronic ground state.

\vspace{1cm}

\bibliography{NVstats_pra.bib}
\bibliographystyle{iopart-num}

\end{document}